\documentclass[letterpaper,10pt]{article}
\usepackage{amsmath, amssymb, graphicx, geometry, color}
\title{Decision dynamics in complex networks subject to mass media and social contact transmission mechanisms}
\author{Carlos Rodr\'{\i}guez Lucatero, Luis Alarc\'on, Roberto Bernal Jaquez$^*$, Alexander Schaum}

\begin{document}

\maketitle

\begin{abstract}
The dynamics of decisions in complex networks is studied within 
a Markov process framework using numerical simulations combined with 
mathematical insight into the process mechanisms. A mathematical discrete-time 
model is derived based on a set of basic assumptions on the convincing 
mechanisms associated to two opinions. The model is analyzed with respect to 
multiplicity of critical points, illustrating in this way the main behavior to 
be expected in the network. Particular interest is focussed on the effect of 
social network and exogenous mass media-based influences on the decision behavior. A set of numerical simulation results is provided illustrating how these mechanisms impact the final decision results. The analysis reveals (i) the presence of fixed-point multiplicity (with a maximum of four different fixed points), multistability, and sensitivity with respect to process parameters, and (ii) that mass media have a strong impact on the decision behavior.
\end{abstract}

\section{Introduction}

Recently, the problem of modeling, analysis and simulation of rumor and decision dynamics in complex networks has obtained increasing attention, including studies on rumor spread in social and small-world networks \cite{Moreno2012Rumor}, and decision dynamics in scale-free networks \cite{Xie1}. While the problem of rumor spreading \cite{Moreno2012Rumor} considers the time evolution of rumor spreaders and stiflers, the decision dynamics \cite{Xie1} is concerned with the time evolution of different decisions in terms of the associated competitive interplay, similar to the one observed in biological studies on different predators competing for the same prey, or chemical species competing for the same reactant. The most notable difference with respect to these classic fields of competition dynamics consists in the fact that in social system dynamics the topology of the underlying contact network is substantial. For example, social phenomena involve different personalities with different number of contacts each, so that the dynamics take place with an intrinsic distributed character and, in particular, emergence of nearly homogeneous groups (so called clusters) is possible. These differences make it necessary to analyze social behavior from the view point of dynamic networks. It is particularly noteworthy that it has been observed that the underlying network topology, unless changing in time, always fulfills the same characteristics in terms of node degree distribution (see e.g. \cite{Barab1, Falou1, Milena1}). For the purpose of understanding the mechanisms underlying dynamic phenomena over dynamic networks, mathematical models can be developed and analyzed formally as well as with numerical simulation studies.  

With particular emphasis on the dynamics of decision competition in scale-free networks, in \cite{Xie1} the regularity of spreading of information and public opinions towards two competing products in complex networks was analyzed. They proposed a simple linear model and simulated the associated decision trajectories over time. It was highlighted that, in contrast to most models studied so far via modified SIS and SIR models \cite{Satorras1, Satorras2, Satorras3, Moreno1, Deepa1, Deepa2, Wan1, Gomez1, CRL1}, in a real life setting there are frequently various information being spread simultaneously (diverse virus, multiple opinions, rumors, etc.), and, in contrast to the single opinion spread dynamics, these may be mutually strenghtened or even annihilated. The competition-based dynamics of two opinions which freely flow in a complex network was studied, showing some interesting and important facts concerning the behavior of decision competition. Nevertheless, studying the underlying mechanisms of mutual 
influence between nodes in the network is expected to lead to a nonlinear dynamics, just as in the SIS and SIR case mentioned above. Accordingly, it may be supposed that the inherent competition mechanisms lead to classical nonlinear phenomena such as multiple attractors and parameter sensitivity.

In the present study a mathematical model is introduced and analyzed for the dynamics of two competing opinions (with neutral intermediate state) which explicitly accounts for the kind of nonlinear interactions inherent to the dynamics of competing opinions in complex social networks. A generalized model for the interaction mechanisms is employed which has been recently proposed in  \cite{Gomez1} and provides intermediate cases between the classical contact and reactive process, and the influence of exogenous mass media opinion propagation into the network is considered explicitly. The mathematical model is analyzed with respect to attractor multiplicity, delimiting in this way some system inherent behavior possibilities. In particular the study focusses on the effect of opinion adaptation through (i) social contacts, and (ii) exogenous mass media influence. In order to analyze these effects, a series of numerical simulations is provided for the two extreme cases of contact and reactive process \cite{Gomez1}, illustrating typical behavior in presence of: 
\begin{itemize}
 \item static contacts
 \item dynamic contacts 
 \item exogenous mass media propagation of one opinion
 \item exogenous mass media propagation of both opinions.
\end{itemize}
The presented results illustrate (i) typical nonlinear behavior such as attractor multiplicity, and sensitivity with respect to process parameters, and (ii) the strong impact mass media have on the decision behavior. The presented results establish an extension of those reported in \cite{Xie1}.

From a methodological point of view, we are putting together the quantitative and qualitative potential of mathematical modeling and simulation, with some basic concepts of dynamical systems theory on attractor multiplicity and stability on the basis of numerical simulation studies.

The paper is organized as follows. In Section 2, the nonlinear opinion dynamics model for dynamic networks is derived based on a set of basic assumptions, and the fixed points are established for the corresponding static and unperturbed version. In Section 3, numerical simulations are presented which illustrate the main behavior observed in decision dynamics in social networks. In Section 4, the main contributions of the study are summarized.

\section{The decision dynamics model}
\newcommand{\A}{\mathcal{A}}
\newcommand{\B}{\mathcal{B}}
\newcommand{\N}{\mathcal{N}}

In this section a mathematical model for the prevalence of two opinions, $\A$ and $\B$, with a neutral intermediate state $\N$ is derived based on the Markov process assumption \cite{Durrett} that the state at any time instant $t\geq1$ ($t\in\mathbb{N}$) depends exclusively on the state at the immediately preceding time instant $t-1$. The result is a discrete time $3N$-dimensional nonlinear model, where $N$ is the number of nodes in the network.

\subsection{Basic assumptions and nomenclature}
\begin{figure}[h]
\centering
\includegraphics[scale=0.4]{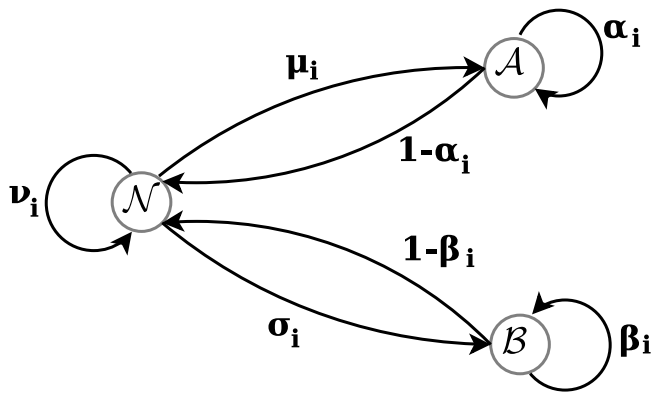}
\caption{State transition diagram.}
\label{StateAutomata}
\end{figure}

In order to derive the model we introduce the following basic assumptions:

\begin{itemize}

\item[(A1)] Two opinions $\A$ and $\B$ are being propagated in a social network with $N$ nodes and Power Law degree ($k$) distribution  \cite{Newman2005}
\begin{align}\label{PowerLaw}
P[k]\sim k^{-\gamma}
\end{align}
(see Figure \ref{Fig:PowerLaw}). The corresponding network topology is reflected in the square adjacency matrix $A(t)$ with time-varying entries $A_{ij}(t)\in\{0,1\}$.

\begin{figure}[!h]
 \centerline{\includegraphics[scale=0.5]{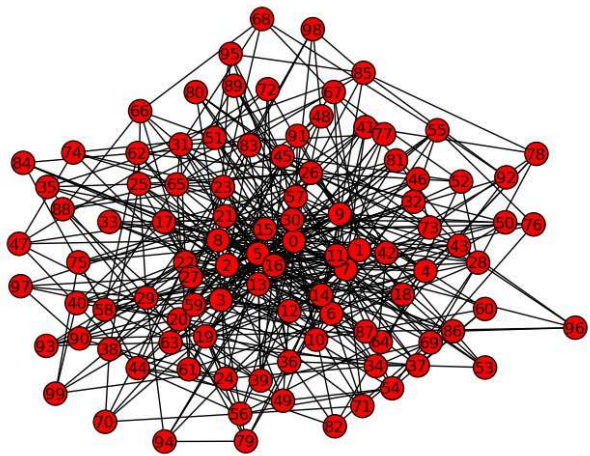}}
 \caption{Power law degree distributed network with $N=100$ nodes, and $P[k]\sim k^{-2.63}$.}
 \label{Fig:PowerLaw}
\end{figure}

\item[(A2)] The total number of nodes $N$ is constant.

\item[(A3)] Any node $i$ can be in one of three states: $\A$ if it has the opinion $\A$, $\B$ if it has the opinion $\B$, and $\N$ if it is neutral with respect to the opinions $\A$ and $\B$. 

\item[(A4)] In order to change from opinion $\A$ to $\B$ (or \textit{vice versa}), the node has to pass through the neutral state $\N$, i.e. there is no direct connection between the states $\A$ and $\B$. The associated state-diagram is depicted in Figure \ref{StateAutomata}.

\item[(A5)] The probability for a node $i$ to be in the state $\A$, $\B$, or $\N$, is denoted by $a_i$, $b_i$ and $n_i$, respectively. It holds that $a_i, b_i, n_i \in[0,1]$ for any $i=1,\ldots,N$.

\item[(A6)] If the node $i$ has the opinion $\A$ (or $\B$), it will convince of opinion $\A$ (or $\B$) the nodes connected with him with probability $\kappa_{A}$ (or $\kappa_B$).

\item[(A7)] A neutral node $i$ (i.e. a node which is in the state $\N$), does not convince any neighbor node.

\item[(A8)] The impact any neighbor node $j\neq i$ has on node $i$ is independent of $j$.

\item[(A9)] The influences of any two nodes $j\neq k\neq i$ on node $i$ are mutually independent. The total effect of all neighbors of node $i$ is given by the mean influence.

\item[(A10)] The opinion $\A$ (or $\B$) is interchanged between two nodes $i$ and $j$ $\lambda_A$ (or $\lambda_B$) times in each time step \cite{Gomez1}. 

\item[(A11)] The mass media influence can be modeled in terms of an exogenous perturbation with period of appearance $T$, taking into account the underlying automata transition possibilities (see assumption A4 and Figure \ref{StateAutomata}).

\end{itemize}

A direct consequence of assumptions A2 and A4 is that
\begin{align}\label{Conservation}
 a_i(t)+b_i(t)+n_i(t) = 1.
\end{align}
Assumption A8 is motivated by the fact that any node $j$ has the same impact on $i$ (assumption A7) and the final state of node $i$ after contacting his neighbors will be a weighted sum of all particular contacts.
Assumption A9 is associated to the type of interchange mechanism \cite{Gomez1}. Two extreme cases have been reported in the literature: the contact process (with $\lambda_k=1,\, k=A,B$), and the reactive process (with $\lambda_k\to\infty,\,k=A,B$). In \cite{Gomez1} a generalized model for the contact rate was proposed covering all intermediate scenarios. In the present study the model proposed in \cite{Gomez1} for the contact rate is adapted by defining the transmission rates associated to opinion $A$ and $B$ as
\begin{align}\label{contacts}
 r_{ij}^A(t) = 1-\left(1-\dfrac{1}{N_i(t)}\right)^{\lambda_A},\quad  r_{ij}^B (t)= 1-\left(1-\dfrac{1}{N_i(t)}\right)^{\lambda_B}
\end{align}
where $N_i(t)$ is the total number of neighbours of node $i$ at time $t$, i,e.
\begin{align}
 N_i(t) = \sum_{j\neq i}A_{i,j}(t).
\end{align}
For the reactive process ($\lambda_k\to\infty,\,k=A,B$) $r_{ij}^k=1,\, k=A,B$. The transition probabilities associated to the process are denoted as follows: 

\begin{itemize}

\item If the node $i$ is in the state $\A$, $\B$ or $\N$, the probability for node $i$ of remaining in $\A$, $\B$, or $\N$ is denoted by $\alpha_i$, $\beta_i$, or $\nu_i$, respectively.

\item If the node $i$ is in the neutral state $\N$, the probability of being convinced of opinion $\A$ (or $\B$) is denoted by $\mu_{i}$ (or $\sigma_i$).

\end{itemize}

\subsection{Derivation of the Markov process model}

Having as point of departure the preceding assumptions, the transition probabilities $\alpha_i, \beta_i, \nu_i, \mu_i$ and $\sigma_i$ are determined. 

As representative case, suppose that a node $i$ is of opinion $\A$ and has contact with a node $j\neq i$. The following events are possible:
\begin{itemize}
 \item The node $j$ is in the state $\A$ (or $\N$) with probability $a_j$ (or $n_j$), and consequently will not convince node $i$ of changing its opinion.
 
 \item The node $j$ is in the state $\B$ with probability $b_j$, and consequently is trying to convince node $i$ with probability of success given by $r_{ij}^B\kappa_B$, and will fail to convince $i$ of opinion $\B$ with probability $1-r_{ij}^B\kappa_B$, with $r_{ij}^B$ given in \eqref{contacts}.
\end{itemize}
On the basis of these considerations, the probability for node $i$ of not being convinced by node $j$ of changing from state $\A$ to $\N$ (the only alternative to remaining in $\A$), is given by the sum
\begin{align}\label{a_ij_first}
 \alpha_{ij} = a_j+n_j+(1-r_{ij}^B\kappa_B)b_j = 1-r_{ij}^B\kappa_Bb_j,
\end{align}
where the last equality is a consequence of property \eqref{Conservation}.

The same reasoning applies to the probability $\beta_{ij}$ of node $i$ being in state $\A$ of not being convinced to change from the state $\B$ to the state $\N$ by its neighbor $j\neq i$ [having in mind property \eqref{Conservation}]:
\begin{align}\label{b_ij_first}
 \beta_{ij} = b_j + n_j + (1-r_{ij}^A\kappa_A)a_j  = 1-r_{ij}^A\kappa_Aa_j.
\end{align}

In the case that node $i$ is in the state $\N$ and in contact with a node $j\neq i$, the following events imply that $i$ remains in $\N$:
\begin{itemize}
 \item The node $j$ is of opinion $\N$ with probability $n_j$, and does not convince $i$ of changing its opinion.
 
 \item The node $j$ is of opinion $\A$ (or $\B$) with probability $a_j$ (or $b_j$, and does convince $i$ of opinion $\A$ (or $\B$) with probability $r_{ij}^A\kappa_A$ (or $r_{ij}^B\kappa_B$), and fail with probability $1-r_{ij}^A\kappa_A$ (or $1-r_{ij}^B\kappa_B$).
\end{itemize}
Accordingly, having in mind property \eqref{Conservation}, the probability for node $i$ of not being convinced by node $j\neq i$ is given by
\begin{align}\label{n_ij_first}
 \nu_{ij} = n_j + (1-r_{ij}^A\kappa_A)a_j + (1-r_{ij}^B\kappa_B)b_j =  1 - r_{ij}^A\kappa_Aa_j-r_{ij}^B\kappa_Bb_j,
\end{align}
and the probabilities of node $j\neq i$ convincing node $i$ of opinion $A$ or opinion $B$ are given respectively by
\begin{align}
 \mu_{ij}  = r_{ij}^A\kappa_A a_j,\quad \sigma_{ij} = r_{ij}^B\kappa_B b_j.
\end{align}
Next, to take into account the combined effect of all neighbors of node $i$ on the transition probabilities, recall assumption A8. The overall transition probabilities for node $i$ at time $t$ are given by:
\begin{align}\label{Transitions}
\begin{array}[c]{l}
 \alpha_i(t) = \dfrac{1}{N_i(t)} \sum_{j\neq i} \alpha_{ij}(t) = \dfrac{1}{N_i(t)} \sum_{j\neq i} [1-r_{ij}^B\kappa_Bb_j(t)],\\
 \beta_i(t) = \dfrac{1}{N_i(t)} \sum_{j\neq i} \beta_{ij}(t) = \dfrac{1}{N_i(t)} \sum_{j\neq i} [1-r_{ij}^A(t)\kappa_Aa_j(t)],\\
 \nu_i(t) =\dfrac{1}{N_i(t)} \sum_{j\neq i} \nu_{ij}(t) = \dfrac{1}{N_i(t)} \sum_{j\neq i} [1-r_{ij}^A(t)\kappa_Aa_j(t)-r_{ij}^B(t)\kappa_Bb_j(t)],\\
 \mu_i(t) = \dfrac{1}{N_i(t)} \sum_{j\neq i} \mu_{ij}(t) = \dfrac{1}{N_i(t)} \sum_{j\neq i} r_{ij}^A(t)\kappa_Aa_j(t),\\
 \sigma_i(t) =\dfrac{1}{N_i(t)} \sum_{j\neq i} \sigma_{ij}(t) = \dfrac{1}{N_i(t)} \sum_{j\neq i} r_{ij}^B(t)\kappa_Bb_j(t).
\end{array}
\end{align}

The associated discrete time model for the dynamics of decisions between $\A$ and $\B$ in the network is given by
\begin{align}\label{Model1}
\begin{array}[c]{l}
 a_i(t+1) = \alpha_i(t)a_i(t)+\mu_i(t)n_i(t),\quad\quad a_i(0)=a_{i0}\\
 n_i(t+1) = \nu_i(t)n_i(t) + [1-\alpha_i(t)]a_i(t)+[1-\beta_i(t)]b_i(t),\quad n_i(0)=n_{i0}\\
 b_i(t+1) = \beta_i(t)b_i(t)+\sigma_i(t)n_i(t),\quad\quad b_i(0)=b_{i0}\\
 0 = 1-[a_i(t)+b_i(t)+n_i(t)].
\end{array}
\end{align}

Taking into account the identities [see \eqref{Transitions}]
\begin{align}
 \mu_i(t) = 1-\beta_i(t),\quad \sigma_i(t) = 1-\alpha_i(t)
\end{align}
model \eqref{Model1} is equivalent to
\begin{align}\label{Model2}
\begin{array}[c]{l}
 a_i(t+1) = \alpha_i(t)a_i(t)+[1-\beta_i(t)][1-a_i(t)-b_i(t)],\quad\quad a_i(0)=a_{i0}\\
 b_i(t+1) = \beta_i(t)b_i(t)+[1-\alpha_i(t)][1-a_i(t)-b_i(t)],\quad\quad b_i(0)=b_{i0}\\
 n_i(t) = 1-a_i(t)-b_i(t)
\end{array}
\end{align}
with $\alpha_i(t)$ and $\beta_i(t)$ defined in \eqref{Transitions}. In vector notation the preceding dynamics are written as
\begin{align}
 x_i(t+1) = f[x_i(t)],\quad x_i(0)=x_{i0},\quad x_i=[a_i,b_i,n_i]\in T,\quad i=1,\ldots, N,
\end{align}
where $T$ is the two-dimensional triangle set (see Figure \ref{Fig:Triangle})
\begin{align}\label{triangleSet}
 T = \{ z=[z_1,z_2,z_3]' \in [0,1]^3\subset\mathbb{R}^3\,| \, z_1+z_2,+z_3=1\}.
\end{align}

Introduce the mean decision probabilities 
\begin{align}
 \rho_A=\dfrac{1}{N} \sum_{i=1}^{\mathcal{N}}a_i,\quad \rho_B=\dfrac{1}{N} \sum_{i=1}^{\mathcal{N}}b_i, \quad \rho_N=\dfrac{1}{N} \sum_{i=1}^{\mathcal{N}}n_i = 1-\rho_A-\rho_B,
\end{align}
and the associated mean probability $\rho(t) = [\rho_a(t),\rho_b(t),\rho_n(t)]'$, it can be easily seen that 
\begin{align}\label{rho}
 \rho(t)\in T,\quad \forall\,t\geq0.
\end{align}

In order to analyze the influence of mass media on the decision behavior, consider the time-varying index subset 
\begin{align}
  I_k(t)\subset\{1,\ldots,N\},\quad \# I_k= [\eta_k N],\quad k=\A,\B,
\end{align}
affecting $\eta\%$ of the total population (here $\# I$ indicates the cardinality of the set $I$), and introduce the associated characteristic function
\begin{align}
 \chi_i^k(t)=\left\{\begin{array}{cc} 1, & i\in I_k(t)\\ 0, & i\notin I_k(t)\end{array}\right.,\quad k=A,B.
\end{align}
In terms of the index subset $I_k,\,k=\A,\B$ the mass media influence can be modeled as follows:
\begin{align}\label{MassMediaModel}
 \begin{array}[c]{l}
    a_i(t+1) = \alpha_i(t)a_i(t)+[1-\beta_i(t)][1-a_i(t)-b_i(t)]+\\
    \hspace{5cm}+\chi_i^A(t)\upsilon^A(t)n_i(t)-\chi_i^B(t)\upsilon^B(t)a_i(t),\quad\quad a_i(0)=a_{i0}\\
    b_i(t+1) = \beta_i(t)b_i(t)+[1-\alpha_i(t)][1-a_i(t)-b_i(t)]+\\
    \hspace{5cm}+\chi_i^B(t)\upsilon^B(t)n_i(t)-\chi_i^A(t)\upsilon^A(t)b_i(t),\quad\quad b_i(0)=b_{i0}\\
 \end{array}
\end{align}
where 
\begin{align}
  \upsilon^k(t)=\left\{\begin{array}{cc}0,& t\neq sT_k,\,s\in\mathbb{N}\\ 1,& t=sT_k,\,s\in\mathbb{N}\end{array}\right.,\quad k=A,B                      
\end{align}
is a function which is nonzero only in discrete time instants $sT_k, s\in\mathbb{N}, k=A,B$ with period $T_k$.

It should be noted that, in comparison to the study presented in \cite{Xie1}, there are some substantial differences to the present work. First, the model built in \cite{Xie1} is linear while the model \eqref{MassMediaModel} is intrinsically nonlinear. Second, the model \eqref{MassMediaModel} allows to take into account different modalities of transmission process between the two extreme cases of contact process ($\lambda=1$) and reactive process ($\lambda\to\infty$) as proposed in \cite{Gomez1}. Third, in \eqref{MassMediaModel} the influence of exogenous opinion propagation into the network is explicitly considered.

\subsection{Basic characterization of the dynamic behavior}

Assume for the moment that the connections between individuals do not vary over time (i.e., the adjacency matrix $A$ is constant, and hence $N_i$ is constant for any node $i$). Based on this assumption the fixed points associated to the dynamics \eqref{Model1} can de determined in order to establish limit case conditions for the time evolution of the opinions $\A$ and $\B$ in the network. 

\begin{figure}[!h]
 \centerline{\includegraphics[scale=0.3]{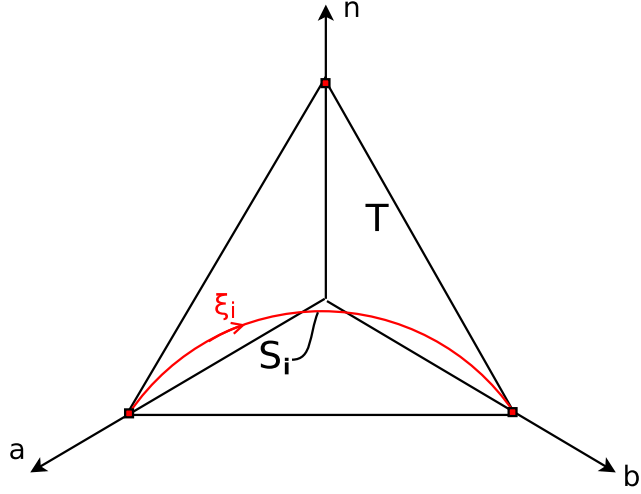}}
 \caption{Sketch of the triangle set $T$ \eqref{triangleSet} where the decision probabilities of each node $i$ and the mean probabilities $\rho$ evolve, and the curve $S$ parameterized by the scalar $\xi$ \eqref{DefXi}.}
 \label{Fig:Triangle}
\end{figure}

The fixed point condition read 
\begin{align}
 a_i(t+1) = a_i(t)=a_i,\quad b_i(t+1)=b_i(t)=b_i,\quad n_i(t+1)=n_i(t)=n_i
\end{align}
where $a_i,b_i, n_i$ are the fixed point values of the stochastic variables $a_i(t),b_i(t)$ and $n_i$, respectively. It follows that 
\begin{align}\label{FixedPoinEqs}
\begin{array}[c]{l}
 0 = (\alpha_i-1)a_i + (1-\beta_i)n_i,\\ 
 0 = (\beta_i-1)b_i + (1-\alpha_i)n_i,\\ 
 0 = -(1-\alpha_i)a_i -(1-\beta_i)b_i+(2-\alpha_i-\beta_i)n_i.
\end{array}
\end{align}
Given that for $a_i=0$ (or $b_i=0$) for all $i$ it holds that $\beta_i=1$ (or $\alpha_i=1$), it follows that the three vertexes 
\begin{align}
 x = \left[\begin{array}{c}1\\0\\0\end{array}\right], \quad  x = \left[\begin{array}{c}0\\1\\0\end{array}\right],\quad  x = \left[\begin{array}{c}0\\0\\1\end{array}\right]
\end{align}
of the triangle set $T$ \eqref{triangleSet} (Figure \ref{Fig:Triangle}) are particular fixed points. Furthermore, solving the equation set \eqref{FixedPoinEqs} in general with respect to $a_i, b_i$ and $n_i$ yields the following set of implicit solutions
\begin{align}\label{FPwithXi}
 a_i = \dfrac{1}{\xi_i^2+\xi_i+1},\quad
 b_i = \dfrac{\xi_i^2}{\xi_i^2+\xi_i+1},\quad 
 n_i = 1-a_i-b_i = \dfrac{\xi_i}{\xi_i^2+\xi_i+1},\quad i=1,\ldots, N,
\end{align}
parameterized by the scalar
\begin{align}\label{DefXi}
\xi_i=\dfrac{1-\alpha_i}{1-\beta_i}\in\mathbb{R},\quad \alpha_i = \dfrac{1}{N_i}\sum_{j\neq i} (1-\kappa_b b_j),\, \beta_i = \dfrac{1}{N_i}\sum_{j\neq i} (1-\kappa_a a_j).
\end{align}
Introducing the solution in vector notation
\begin{align}
 s_i = [a_i,b_i,n_i]'\in T,
\end{align}
the following limit cases can be directly derived: 
\begin{align}\label{LimitCases1}
\kappa_A\to0,\,\Rightarrow\, \xi_i\to\infty\,\text{ and }\,s_i\to \left[\begin{array}{c}1\\0\\0\end{array}\right],\quad \kappa_B\to0\,\Rightarrow\, \xi_i\to 0\,\text{ and }\, s_i\to\left[\begin{array}{c}0\\1\\0\end{array}\right],
\end{align}
corresponding to the two bottom vertexes of the triangle set $T$ \eqref{triangleSet} (Figure \ref{Fig:Triangle}). In terms of the opinion density $\rho$ \eqref{rho} there are four associated fixed points
\begin{align}
 \rho \in \left\{\left[\begin{array}{c}1\\0\\0\end{array}\right], \left[\begin{array}{c}0\\1\\0\end{array}\right], \left[\begin{array}{c}0\\0\\1\end{array}\right], \dfrac{1}{N}\left[\begin{array}{c}\sum_{i=1}^Na_i\\\sum_{i=1}^Nb_i\\\sum_{i=1}^Nn_i\end{array}\right]\right\},
\end{align}
where $a_i,b_i$ and $n_i$ are given in \eqref{FPwithXi}. Naturally, the limit cases \eqref{LimitCases1} can be expressed in terms of the mean probability solution $s_\rho$ as
\begin{align}\label{LimitCases2}
 \kappa_A\to0,\,\Rightarrow\, s_\rho \to \left[\begin{array}{c}1\\0\\0\end{array}\right],\quad \kappa_B\to0\,\Rightarrow\, s_\rho\to  \left[\begin{array}{c}0\\1\\0\end{array}\right],\quad s_\rho = \dfrac{1}{N}\sum_{j\neq i}s_i.
\end{align}

Summarizing, for any parameter combination $(\kappa_A,\kappa_B,\lambda_A,\lambda_B)>0$ there are four fixed points, and in the limit $\kappa_A=0$ (or $\kappa_B=0$) there are three fixed points. This fact illustrates the multiplicity of fixed points and sensitivity with respect to the dynamic's parameters, and reveals that the nonlinearity introduced by the interchange mechanisms is too strong to be neglected. This fact establishes a main difference with respect to the recent study \cite{Xie1}, where a linear model was used for approximating the decision dynamics. A formal analysis of the stability properties of the four fixed points nevertheless goes beyond the scope of the present study, and here we circumscribe ourselves to numerical simulation studies in order to find the main behavior expected in decision dynamics in social networks, and study the impact of mass media exogenous perturbations.

\section{Simulation results}
In this section, the dynamic behavior of decisions in social networks is analyzed for four different scenarios:
\begin{itemize}
 \item Static network, i.e. static adjacency matrix with power law (scale free) distribution.
 \item Dynamic network, i.e. time-varying adjacency matrix with power law distribution over any time interval.
 \item Dynamic network with exogenous mass media influence of one opinion on 18 $\%$ of the population with period $T_A = 8$ time units.
 \item Dynamic network with exogenous mass media influence of both opinion on 10 $\%$ of the population with period $T_A = 8$ time units and $T_B = 16$ time units.
\end{itemize}
The simulations were taken out for a total population of $N=10.000$ nodes, and a power law node degree $k$ distribution \eqref{PowerLaw} with \cite{Newman2005}
\begin{align}
 \gamma = 2.16.
\end{align}

\subsection{Static network without exogenous perturbations}

In order to illustrate the main (nonlinear) behavior of the decision dynamics process according to the fixed point multiplicity discussed in Section 2.3, the two limit cases with $\lambda_k  = 1, k=A,B$ (contact process) and $\lambda_k\to\infty, k=A,B$ (reactive process) are illustrated for three different parameter scenarios. For the purpose at hand the projection of the trajectories in the triangle set $T$ \eqref{triangleSet} onto the $(\rho_a,\rho_b$)-plane is presented. Note that accordingly it is possible that trajectories seem to intersect in the projection.

\begin{figure}[!h]
 \centerline{\includegraphics[scale=.7]{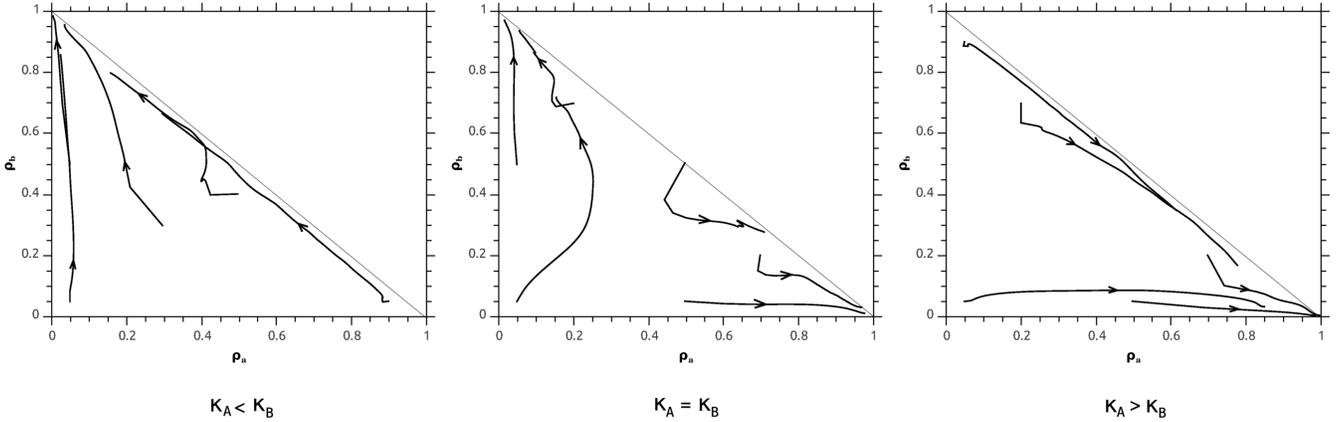}}
 \caption{Projection onto the $(\rho_a,\rho_b)$-plane of the trajectories of the mean decision probability $\rho$ \eqref{rho} for the three different parameter sets given in \eqref{ParamSet1} and a contact transmission process.}
 \label{CP_static}
\end{figure}

For the purpose at hand consider the following three parameter sets:
\begin{align}\label{ParamSet1}
  \kappa_A = 0.1,\kappa_B = 0.9 \,(\kappa_A<\kappa_B),\quad \kappa_A  =\kappa_B = 0.5\,(\kappa_A=\kappa_B),\quad \kappa_A = 0.9,\kappa_B=0.1\,(\kappa_A>\kappa_B)
\end{align}
and that $\A$ and $\B$ are propagated via contact processes (i.e., $\lambda_A=\lambda_B=1$). The corresponding simulation results are presented in Figure \ref{CP_static}.

It can be seen in Figure \ref{CP_static} that: 
\begin{itemize}
 \item[(i)] for the first parameter set $(\kappa_A<\kappa_B$) all the trajectories move towards an attractor in the upper left corner of the ($\rho_A,\rho_B$)-plane.

 \item[(ii)] for the second parameter set $(\kappa_A=\kappa_B$) there are two attractors (one in the upper left and one in the lower right corner) towards which the decision trajectory $\rho$ may converge, in dependence on the initial value.

 \item[(iii)] for the third parameter set $(\kappa_A>\kappa_B$) all the trajectories move towards an attractor in the lower right corner of the ($\rho_A,\rho_B$)-plane.
\end{itemize}
According to these results, there is a strong sensitivity, in the kind of structural instability \cite{Andronov1937}, of the decision behavior in the network on the convincing parameter pair ($\kappa_A, \kappa_B$), manifesting itself in particular through the fact that, in the passage from the first to third parameter set in \eqref{ParamSet1}, the attractor at the lower right corner initially is unique, than coexists with a second attractor in the upper left corner, and finally dissapears leaving the attractor in the upper left corner as the unique one. Hence, for some parameter combinations the final decision state will depend strongly on the initial condition.

It should be noticed that close to the fixed-point, the dynamics becomes very slow in correspondence to the fact that almost the whole network is of one single opinion, so that the existing contacts only confirm the present decision state. This behavior is predicted by the model \eqref{Model2}, given that close to the fixed point the functions $\alpha_i,\beta_i\approx 1$ for any node $i$, and hence the system attains a slowly varying (i.e., quasi steady-state) behavior according to
\begin{align}
   a_i(t+1)\approx a_i(t),\quad b_i(t+1)\approx b_i(t),\quad n_i(t+1)\approx n_i(t).  
\end{align}
Our conjecture is that this fact is also related to the generation of decision clusters within the network.
  
\begin{figure}[!h]
 \centerline{\includegraphics[scale=0.7]{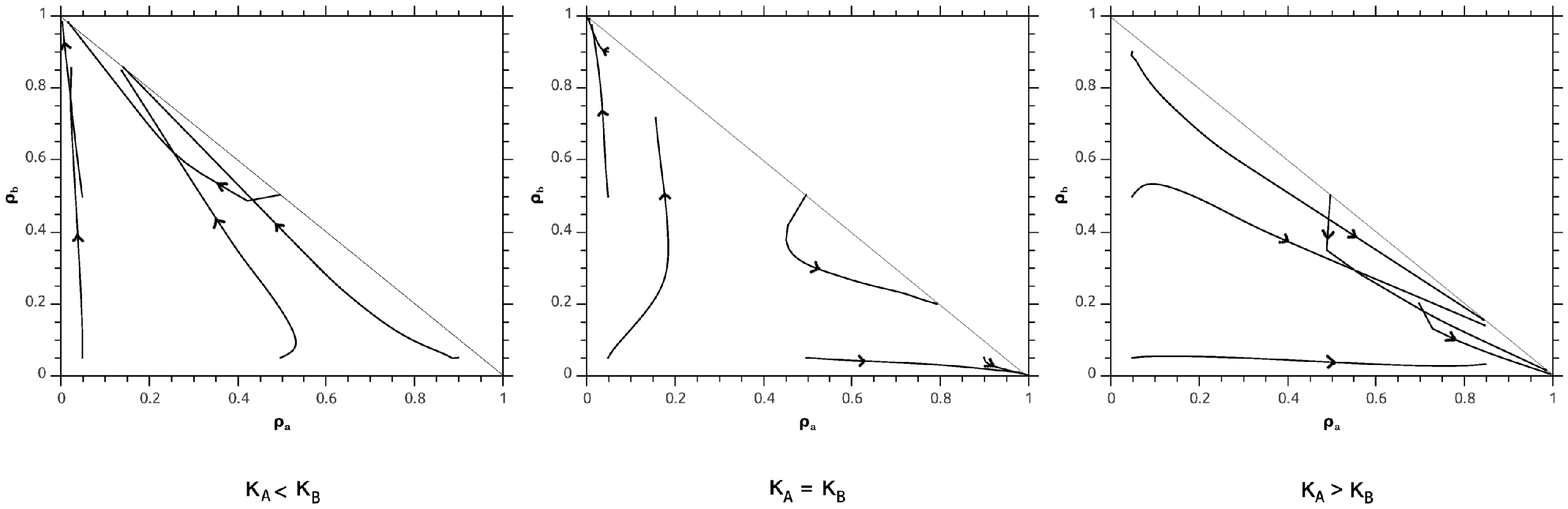}}
 \caption{Proyection onto the $(\rho_a,\rho_b)$-plane of the trajectories of the mean decision probability $\rho$ \eqref{rho} for the three different parameter sets given in \eqref{ParamSet1} and a reactive transmission process.}
 \label{RP_static}
\end{figure}

In Figure \ref{RP_static} are presented the simulation results for the three parameter combinations defined in  \eqref{ParamSet1}, but for reactive processes with $\lambda_A=\lambda_B\to\infty$, or equivalently $r_{ij}^A=r_{ij}^B=1$. It can be observed that the global over-all behavior is quite similar to the one observed for the contact process. Comparing the behavior over time, the only substantial difference consists in that the convergence speed is considerably faster for the reactive process, in accordance to the fact that the transmission rate is much higher.

\subsection{Dynamic network without exogenous perturbations}

Next, consider the case that the network connections vary with time in such a way that the overall degree distribution \eqref{PowerLaw} is maintained. In this case the system dynamics \eqref{Model2} are no longer autonomous, and thus no fixed points exist for the system (unless in the mean, i.e. in terms of $\rho_k, k=A,B,N$, asymptotical stability-like behavior may be observed). 

\begin{figure}[!h]
 \centerline{\includegraphics[scale = 0.8]{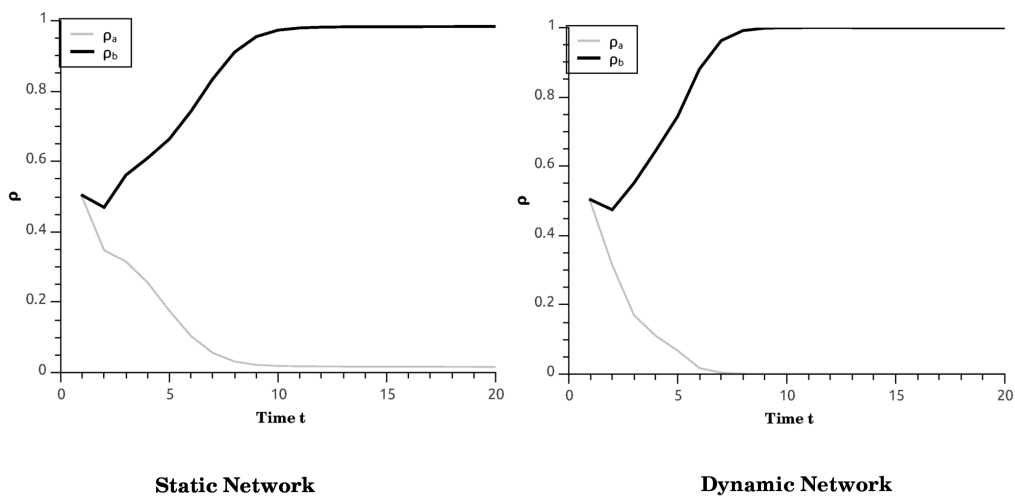}}
 \caption{Time evolution of the mean decision trajectories associated to decision $\A$ (thin grey line) and $\B$ (thick black line) for the parameter value and initial condition set \eqref{Ini0Dynamic} and a reactive transmission process in a scale-free dynamic network.}
 \label{DNRP}
\end{figure}

The associated decision time evolution is compared in Figure \ref{DNRP} to the static network case for the parameter value and initial condition set
\begin{align}\label{Ini0Dynamic}
 \kappa_A=0.1,\quad\kappa_B=0.9,\quad \rho_a(0)=\rho_b(0)=0.5,\quad \rho_n(0)=0
\end{align}
and a reactive process (i.e., $r_{ij}^k=1, k=A,B$). It can be observed that the over-all behavior is quite similar, with the important difference that for the dynamic network case the trajectories are smoother and converge in about $9$ time steps, while in the static network set-up still the fixed-point is not reached. These observations are in line with the results presented in \cite{Schwarzkopf1}, and are probably a result of the fact that in the dynamic network set-up, i.e. with contacts changing at each time instant, the formation of decision clusters is less probable than in the static network set-up. This conjecture should be carefully studied in future research.

\subsection{Dynamic network with mass media influence of one single opinion}

\begin{figure}[!h]
 \centerline{\includegraphics[scale=0.7]{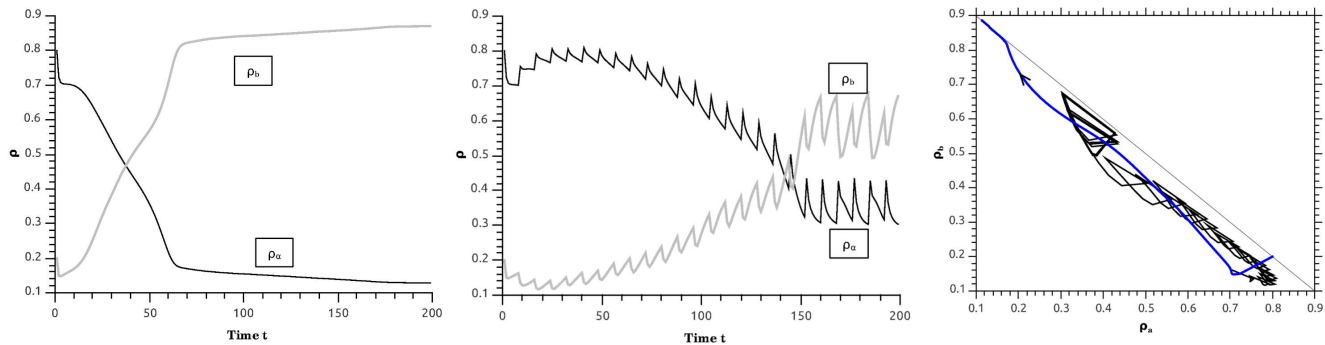}}
 \caption{Left: Time evolution of the unperturbed decision dynamics on a dynamic network for the parameter values and initial conditions given in \eqref{Ini1Propag} and a reactive transmission process. Center: Time-evolution of the associated decision dynamics with exogenous influence on $18\%$ of the population each $T_A=8$ time units. Right: Projection onto the $(\rho_a,\rho_b)$-plane of the trajectories of the mean decision probability $\rho$ \eqref{rho} for the unperturbed dynamics (thick blue line), and the one with exogenous perturbation (black line).}
 \label{Fig:PropagA}
\end{figure}

To analyze the impact of mass media on the decision behavior of a dynamic network, the case of exogenous propagation of opinion $\A$ is studied based on the model \eqref{MassMediaModel} with 
\begin{align}
 T_A = 8 \text{ time units }, T_B= \infty.
\end{align}
In Figure \ref{Fig:PropagA} are presented the simulation results for the parameter values and initial conditions
\begin{align}\label{Ini1Propag}
 \kappa_A=0.3,\,\kappa_B=0.6,\quad \rho_A(0)=0.8,\quad \rho_B(0)=0.2,\quad \rho_N(0)=0. 
\end{align}
If there were no external forces the trajectory would converge towards the upper left corner corresponding to decision $\B$, as can be seen in the corresponding time-evolution shown in the left sub-figure of Figure \ref{Fig:PropagA}, and in the projection onto the $(\rho_a,\rho_b$)-plane shown in the right sub-figure by the thick blue line. For the case that $\A$ is propagated through an exogenous force with arbitrary $18\%$ of the population being impacted each $T_A=8$ time units the corresponding time evolution is shown in the central sub-figure of Figure \ref{Fig:PropagA}, and the projection onto the ($\rho_a,\rho_b$)-plane is shown in the right sub-figure by the black line. It can be seen that at each period there is a force pushing $18\%$ of the population towards decision $\A$. Accordingly, the trajectory does not reach the attractor (corresponding to the decision $\B$), but is maintained in an oscillatory regime about some intermediate decision distribution. This illustrates the impact of exogenous forces on the dynamic behavior in the network and the influence on the overall (mean) decision vector $\rho$.

\subsection{Dynamic network with mass media influence of both opinions}

\begin{figure}[!h]
 \centerline{\includegraphics[scale=0.7]{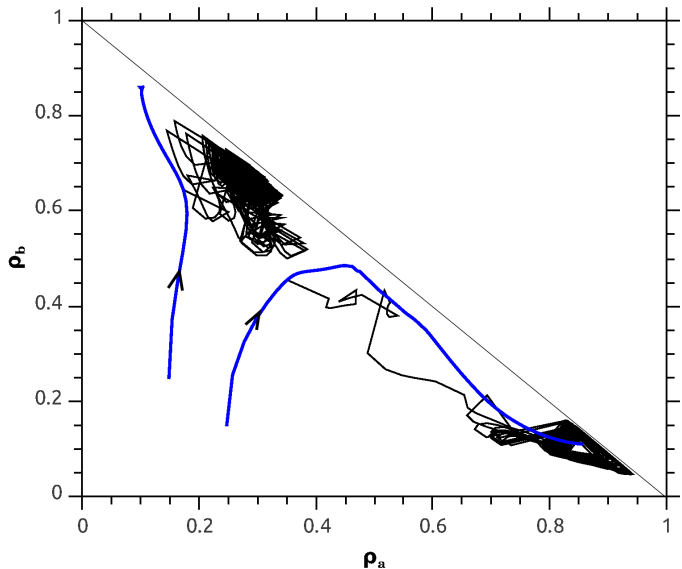}}
 \caption{Projection onto the $(\rho_a,\rho_b)$-plane of two representative trajectories of the mean decision probability $\rho$ \eqref{rho} for the parameter values and initial conditions given in \eqref{Ini2Propag} and a reactive transmission process. The unperturbed trajectories are represented by the thick blue lines, and the ones with exogenous perturbation by the black lines.}
 \label{Fig:PropagAB}
\end{figure}

In order to illustrate the competition of two external forces, consider the case that both decisions are propagated through external forces into the network (e.g. by exogenous mass media) but with different periods
\begin{align}\label{PeriodsAB}
 T_A = 8\text{ time units}, \quad T_B = 16 \text{ time units}.
\end{align}
It is considered that both mechanisms affect $10\%$ of the total population to their favor. The parameter values and initial conditions were set to
\begin{align}\label{Ini2Propag}
 \kappa_A=\kappa_B=0.5,\quad \rho_{1}(0)=[0.25,0.15,0.6]',\quad \rho_{2}(0)=[0.15,0.25,0.6]'
\end{align}
corresponding to two representative cases as shown in Figure \ref{Fig:PropagAB}. 
As illustrated by the thick blue lines, in the unperturbed case, the trajectory starting at $\rho_{1}(0)$ \eqref{Ini2Propag} converges towards the attractor at the upper left corner, and the trajectory starting at $\rho_{2}(0)$ \eqref{Ini2Propag} converges towards the attractor at the lower rightt corner. In presence of the exogenous perturbations with periods given in \eqref{PeriodsAB} and affecting both a total number of $10\%$ of the population to their favor, the trajectories are given by the black lines. It can be seen that the trajectory starting at $\rho_{1}(0)$ does no longer converge towards the attractor at the upper left corner, but enters into an oscillatory regime about some intermediate mean decision value $\rho$, and the trajectory starting at $\rho_{2}(0)$ almost converges towards the lower right corner but enters into an oscillatory regime close to the corresponding attractor for the unperturbed trajectory. 

These results illustrate the differences that may be caused by exogenous propagation mechanisms on the decision dynamics in complex networks. In particular it reveals that the final behavior still represents nonlinear character implying that the final decision state will strongly depend on the initial condition.

\section{Concluding remarks}

The decision dynamics in a scale-free network has been analyzed with respect to inherent nonlinear convincing mechanisms, impact of network variations over time, and mass media favoritism of one single or both opinions. A mathematical model was derived on the basis of some assumptions establishing a quantitative means to analyze the main mechanisms present in decision dynamics. Based on a preliminary analysis of the system's fixed points, it was shown that for a static network there are between three and four fixed points, as a consequence of the system nonlinearity. Numerical simulation studies were presented illustrating the predicted behavior for static and dynamics scale-free networks, which is in correspondence to the one expected in real networks. The impact that mass media have on the decision dynamics was illustrated in different characteristic situations.

\bibliographystyle{unsrt}
\bibliography{carlos}

\end{document}